\documentclass[useAMS, usenatbib]{mn2e}

\usepackage[dvips]{graphicx} 
\input{epsf}
\title{Spectroscopic Identifications of SWIRE sources in ELAIS-N1}

\author[M. Trichas et al.]
         {M. Trichas$^1$\thanks{Email: m.trichas@imperial.ac.uk},
M. Rowan-Robinson$^1$, 
A. Georgakakis$^2$,
I. Valtchanov$^3$,
K. Nandra$^1$, \newauthor
D. Farrah$^4$,
G. Morrison$^{5,6}$,
D. Clements$^1$,
I. Waddington$^4$
\\
        $^1$Astrophysics Group, Blackett Laboratory, Imperial College London,
        Prince Consort Road, London SW7 2BW, UK.\\
        $^2$National Observatory of Athens, I. Metaxa $\&$ V. Pavlou, Athens 15236, Greece\\
        $^3$Herschel Science Centre, ESAC, ESA, Spain\\
        $^4$Astronomy Centre, Department of Physics \& Astronomy, 
        University of Sussex, Brighton, BN1 9QH, UK.\\
        $^5$Institute for Astronomy, University of Hawaii, Honolulu, HI, 96822, USA.\\
        $^6$Canada-France-Hawaii Telescope, Kamuela, HI, 96743, USA}
          \date{Accepted in MNRAS.  }
\pagerange{\pageref{firstpage}--\pageref{lastpage}}
\pubyear{2009}

\begin{document}
\maketitle
\label{firstpage}
\newcommand{\mic}{\umu m}
\newcommand{\gsim}{\mathrel{\rlap{\lower4pt\hbox{\hskip1pt$\sim$}}
    \raise1pt\hbox{$>$}}}   

\pagerange{\pageref{firstpage}--\pageref{lastpage}}
\pubyear{2009}
\begin{abstract}
We present the largest spectroscopic follow-up performed in SWIRE ELAIS-N1. We were able 
to determine redshifts for 289 extragalactic sources. The values of 
spectroscopic redshifts of the latter have been compared with the estimated values from our photometric redshift 
code with very good agreement between the two for both galaxies and quasars. Six of the quasars are hyperluminous
infrared galaxies all of which are broad line AGN. We have performed emission line diagnostics for 30 sources using the $[OIII]$/$H$$\beta$,  $[NII]$/$H$$\alpha$ and $[SII]$/$H$$\alpha$ line ratios in order to classify these 30 sources into star-forming, Seyferts, composite and LINER and compare the results to the predictions from our SED template fitting methods and mid-IR selection methods.
 \end{abstract}

\newcommand{\mnras}{MNRAS}
\newcommand{\apj}{ApJ}
\newcommand{\apjl}{ApJL}
\newcommand{\apjs}{ApJS}
\newcommand{\aj}{AJ}
\newcommand{\aap}{AAP}
\newcommand{\araa}{ARA\&A}
\newcommand{\pasp}{PASP}
\newcommand{\nat}{Nature}

\def\lsim{\mathrel{\rlap{\lower4pt\hbox{\hskip1pt$\sim$}}
    \raise1pt\hbox{$<$}}}     

\section{Introduction}
\begin{figure*}
\begin{center}
\includegraphics[width=8.8 cm]{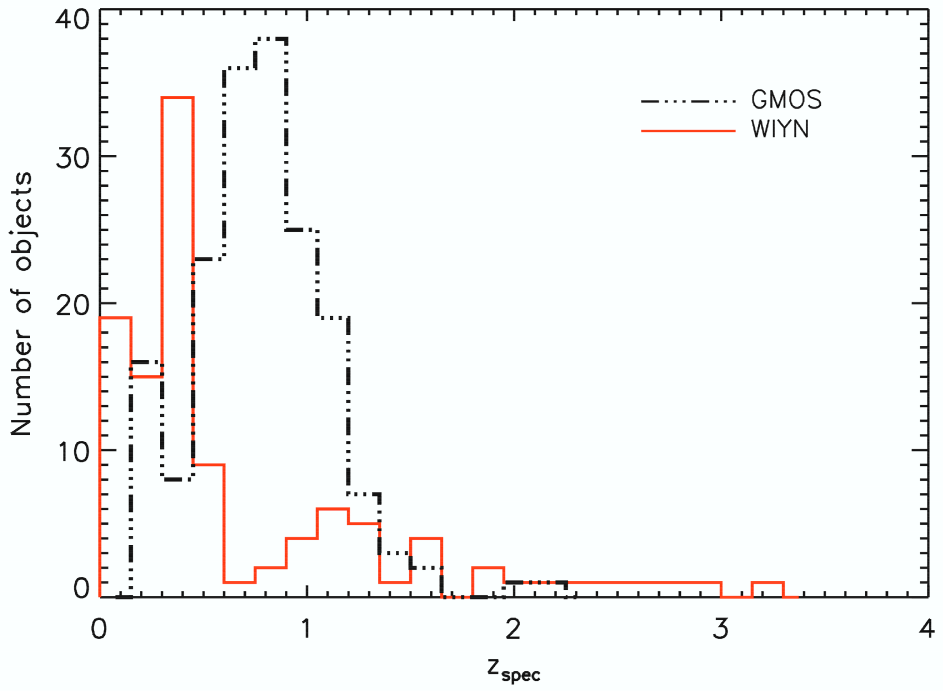}
\includegraphics[width=8.2 cm]{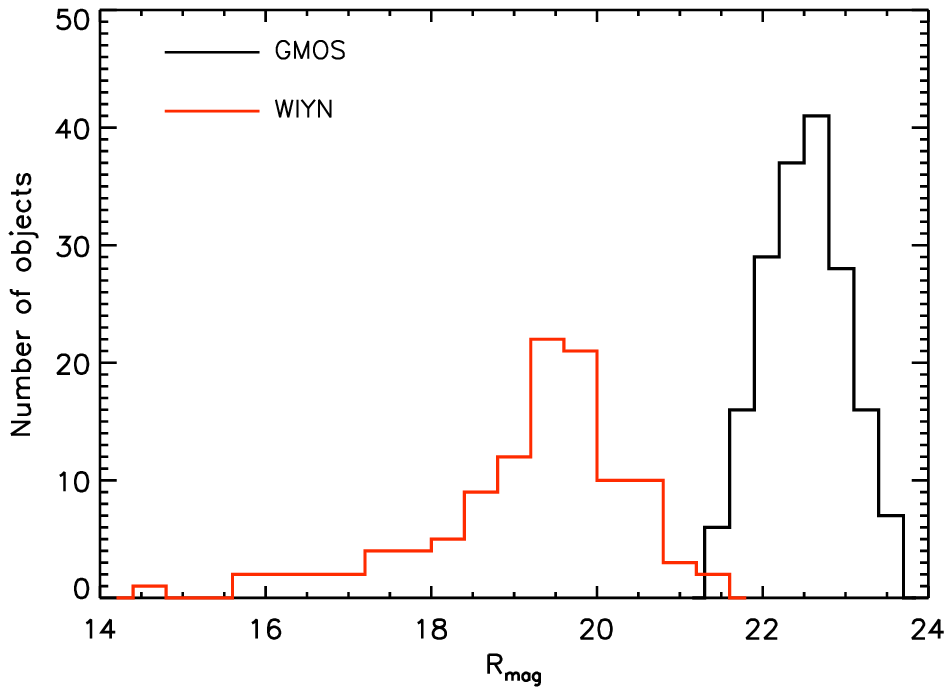}
\caption{$\bf{Left:}$The overall redshift distribution of all extragalactic sources in the final spectroscopic catalogue, 
divided into objects with spectroscopy from Gemini GMOS (dashed black line) and from WIYN Hydra (solid red line). 
$\bf{Right:}$ R-band distribution for the extragalactic sources.  Lines and colours same as left.}
\end{center}
\end{figure*}A major challenge in the quest to constrain the accretion history of the Universe is to compile unbiased samples of AGN. Obscuration of the AGN emission introduces severe selection effects against certain types of AGN. Although X-ray observations, and especially hard X-rays ($2-10$ keV), have been shown to be very efficient in minimizing these biases, in the presence of large obscuring gas column densities ($>$$10^{24}$ $cm^{-2}$) hard X-ray surveys are not able to fully sample the entire population of obscured AGN (Georgantopoulos et al., 2007). \\
The bias of hard X-ray imaging surveys against heavily obscured AGN has motivated alternative methods for finding such systems, especially at mid--infrared wavelengths. The disadvantage of mid-IR surveys is that the AGN are outnumbered by star-forming galaxies and some selection method is necessary to separate the two populations. It has been suggested that mid-infrared  colour-colour plots can be used to effectively isolate obscured AGN. The principle behind this approach is that luminous AGN are characterized by red and almost featureless spectra in mid-infrared. These properties make their IRAC/MIPS colors unique among other infrared sources, such as galaxies and stars, providing a potentially powerful tool to identify them. Different combinations of mid-infrared colors have been proposed to select AGN. Lacy et al. (2004) and Hatziminaoglou et al. (2005) define their selection criteria based on the mid-infrared colours of luminous high-z QSOs identified in the SDSS. A large fraction of the sources defined as AGN in this way have no X-ray counterparts, tentatively suggesting obscuration of the central engine. Although, these selection methods have been successful in selecting obscured AGN, the number
found does not match the predictions of some models for the X-ray background (Georgantopoulos et al. 2008). On the other hand, good quality optical spectroscopic data is a powerful tool for finding AGN even if they are heavily obscurated and low luminosity (Kauffmann et al. 2004). \\
The motivation behind this paper is to test the AGN selection methods in infrared by combining infrared data with optical spectroscopic data. 
In this paper we present results of the largest spectroscopic survey in the ELAIS-N1 region of the Spitzer-SWIRE survey
(Lonsdale et al 2003, 2004).  Section 2 describes the observations, section 3 analyzes the
results, discussing hyperluminous infrared galaxies in the sample and the comparison of photometric and spectroscopic redshifts.
Section 4 discusses emission-line diagnostics and section 5 discusses infrared colour-colour diagrams and spectral energy
distributions.  Section 6 gives a summary of our results. A cosmological model with $\Omega_{o}~=~0.3$, $\lambda_{o}~=~0.7$ and a Hubble constant of 72 $Km~s^{-1}~Mpc^{-1}$ is used throughout.
\section{Observations}
\subsection{Sample selection}
The spectroscopic targets for the Gemini North and WIYN runs were
selected from the available multi-wavelength data in SWIRE-EN1 field.\\
Target lists were prepared from the then latest version of the SWIRE Photometric Redshift Catalogue
(Rowan-Robinson et al 2005), selecting sources with $21.5<r<23.5$~mag
for Gemini and $r<21$ for WIYN.  For Gemini, the candidate sources were
distributed between three prioritized groups, with top priority given to
X-ray sources in the {\it Chandra\/} fields. Next priority were sources
with either (i) optical or infrared SEDs identified as AGN in the
catalogue,  (ii) photometric redshifts $z>1$.  Finally, all other
SWIRE sources with $21.5<r<23.5$ were assigned to the lowest priority
group (iii). For WIYN, the following priority was used: (i) brightest cluster
galaxies (see below), (ii) X--ray detected sources (from both {\it Chandra}
and {\it XMM-Newton}), (iii) AGN selected infrared sources, (iv) galaxies from the
region between the two clusters, (v)  24$\mu m$ sources. Table 1 gives the number of the sources observed per priority list.
\subsection{Multi-wavelngth data} 
$\bf{Optical/HST:}$ there is a single
deep {\it Hubble} Space Telescope observation within the SWIRE-EN1
field (Benitez et al 2004), taken as part of the early release
observations of the Advanced Camera for Surveys (ACS). The ACS field is
centred on the Tadpole galaxy (also known as UGC~10214, VV~29 and
Arp~188), a bright spiral at $z=0.032$ with an extended tidal tail
(Jarrett et al 2006).  This 14~square arcmin image includes 156 galaxies
detected by both ACS and SWIRE (Hatziminaoglou et al 2005).\\
$\bf{Optical/ Infrared:}$ The optical and infrared observations are
part of the Spitzer-SWIRE survey (Lonsdale et al 2003, 2004). \\
$\bf{X-ray:}$ Within the SWIRE-EN1 field, we have X-ray observations
taken as part of the ELAIS Deep X-ray Survey (Manners et al 2003). 
These consist of a single 71.5-ks pointing with the {\it Chandra\/}
X-ray observatory's Advanced CCD Imaging Spectrometer, covering an area
of $17\times17$~square arcmin.  There are 102 sources in common between
{\it Chandra\/} and SWIRE, with X-ray fluxes brighter than
$2.3\times10^{-15}$~ergs~s$^{-1}$~cm$^{-2}$ in the 0.5--8~keV band
(Franceschini et al 2005). There are also a number of {\it XMM-Newton} observations (PI R. Mann)
but unfortunately only three are good enough and not heavily affected by
periods of high background. These three observations with the same
centre cover a field of $\sim 15\arcmin$ radius, south of the region
covered by {\it Chandra}. Two serendipitous galaxy clusters, separated
by $13\arcmin$, were found in the XMM field (Valtchanov et al. in
preparation) and the brightest cluster galaxies as well as some other
putative cluster members were included in the target lists.
\subsection{Gemini}
We have observed the SWIRE ELAIS-N1 field with the Gemini Multi-Object
Spectrogaph (GMOS) on the 8.1-m Gemini North telescope.  The
observations were carried out in queue-scheduled mode during 2005 April
03 to 2005 May 15.  GMOS provides multi-slit optical spectroscopy over a
5.5-arcmin field of view.  We used the 400-lines/mm red grating
(R400\_G5305) at a central wavelength of 7500~\AA.  Nod-and-shuffle mode
was used, with a 3.0~arcsec nod along a 5-arcsec slit and a charge
shuffle of 5.1~arcsec.  A single integration consisted of fifteen
cycles, each cycle being a sequence of four nods in a BAAB configuration
(where A \& B represent the different nod positions), each of 30~s
duration, giving an integration time of 1800~s.  This was then repeated
with the central wavelength offset by 50~\AA, to give full wavelength
coverage across the gaps between different CCDs. Thus each source was
observed with a total integration time of 3600~s.\\
The mask preparation software used the target list priorities
automatically assigning slits to objects in the target list and we were
able to place 15--20 slits on each of the thirteen masks (excluding
acquisition/guide stars), giving 230 objects in total.\\
The HST/ASC field was covered with one GMOS pointing. The {\it Chandra}
field was covered with six pointings of GMOS, each pointing having a
field of view of $5.5\times5.5$~square arcmin, observing a total of 14
X-ray AGN from the ELAIS N1 Deep X--ray survey. In addition, we selected
a further six GMOS pointings, one in the XMM field and the rest randomly
scattered across the SWIRE field.
\begin{table}
\caption{Number of sources observed per priority group. Details of each priority group are given in $\S$2.1.}
\begin{tabular}{lll}
\hline
\hline
Priorities & GMOS & WIYN\\
\hline
\hline
i & 45 & 23 \\
ii & 101 & 34 \\
iii & 63 & 48 \\
iv&-& 12\\
v&-&50\\
\hline
total&209&167\\
\hline
\hline
\end{tabular}
\end{table}
\subsection{WIYN} A selected number of sources from an area of
approximately one square degree was included for spectroscopic
observations on the NOAO WIYN 3.5m telescope using the HYDRA
multi--object bench spectrograph. The observations were performed with
the "red" and "blue" wing of the camera on two different runs: the "red"
run took place on 12-14 April 2005 and the "blue" run on 2--4 July 2005.\\
An automated procedure allocated the fibers in the field. In a couple of
cases an optimization by hand was possible. For the second run (using
the blue wing of the spectrograph) the priority was given to those
objects already observed with the red spectrograph and the new objects
were concatenated at the end of the list.
\begin{figure*} 
\begin{center}
\includegraphics[width=7. cm]{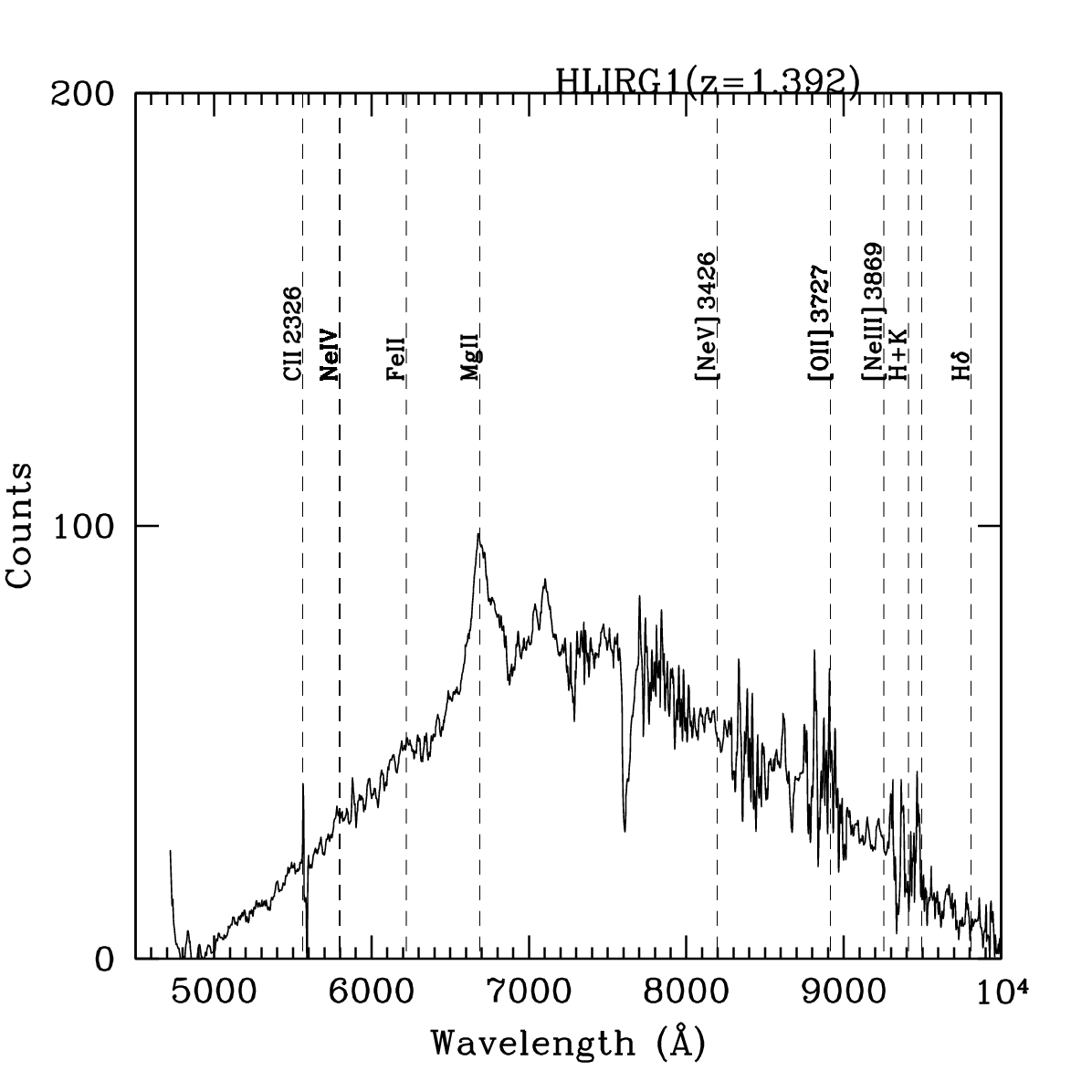}
\includegraphics[width=7. cm]{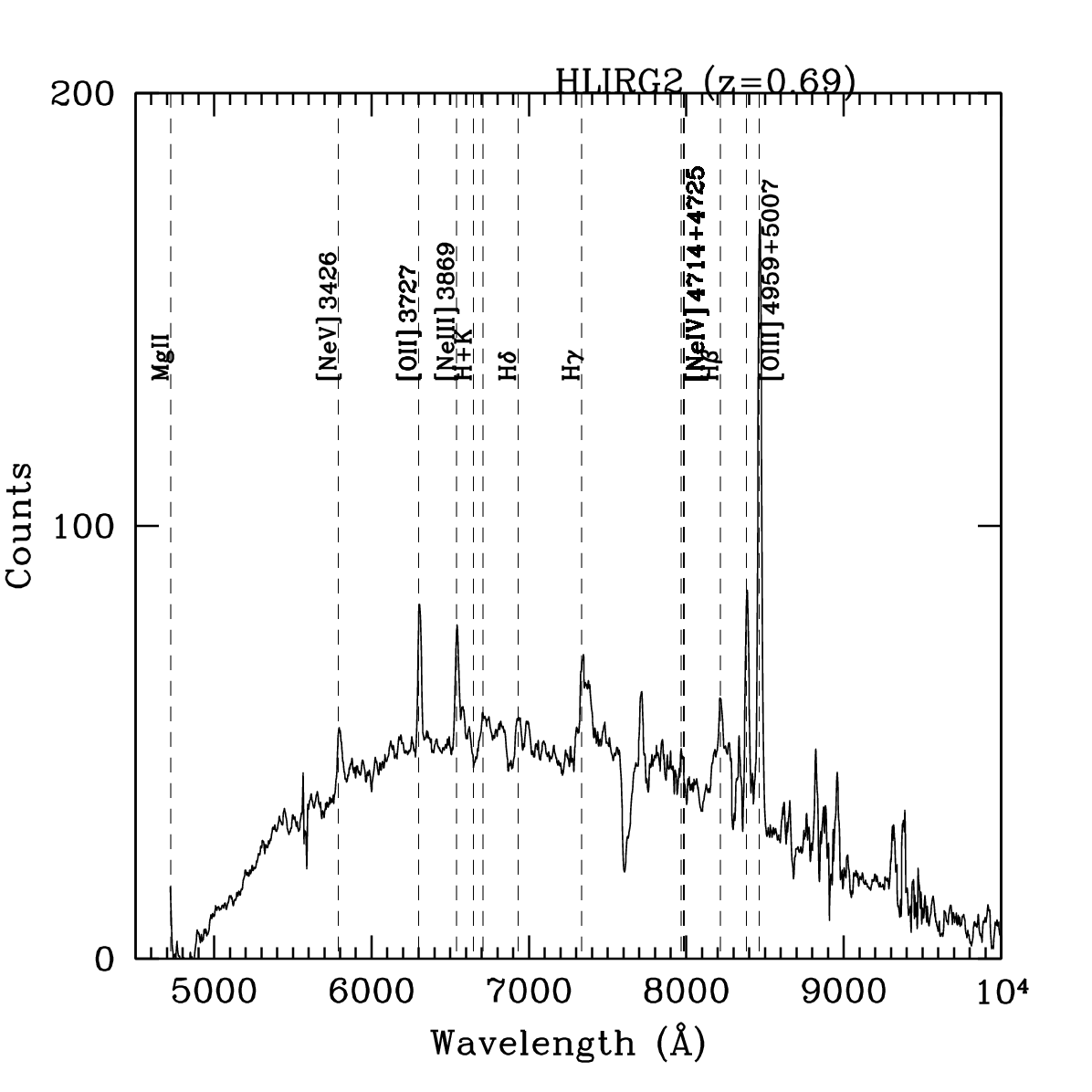}
\includegraphics[width=7. cm]{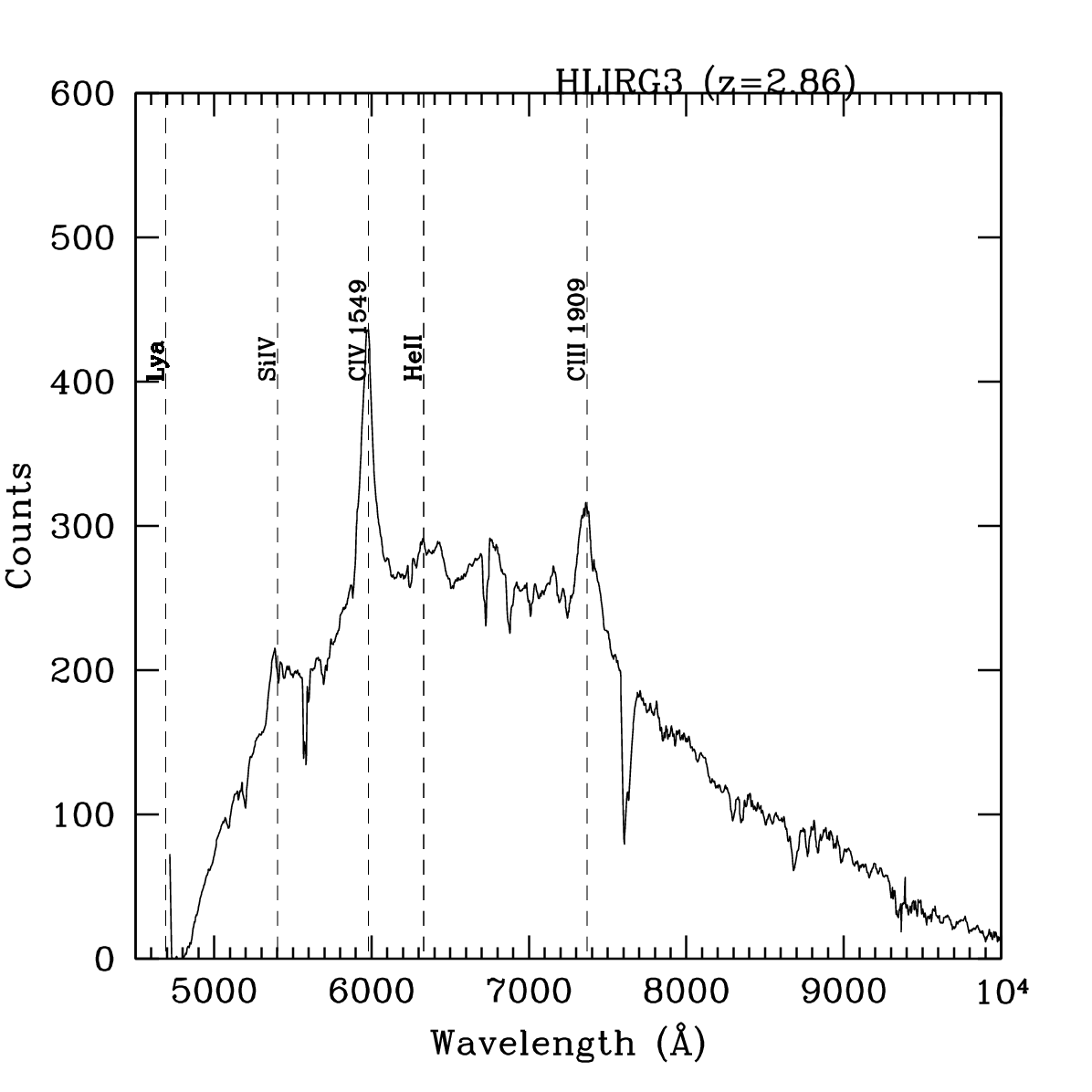}
\includegraphics[width=7. cm]{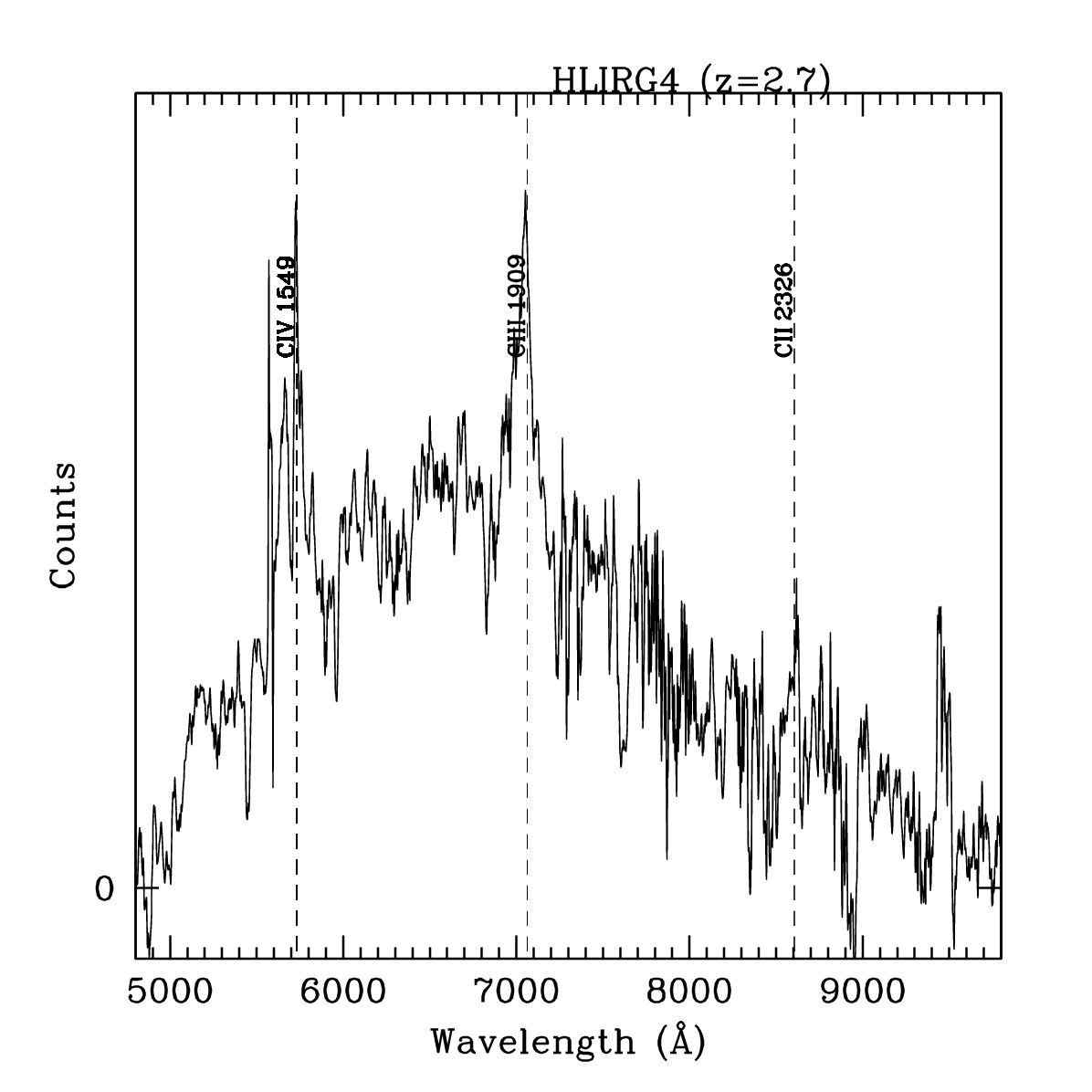}
\includegraphics[width=7. cm]{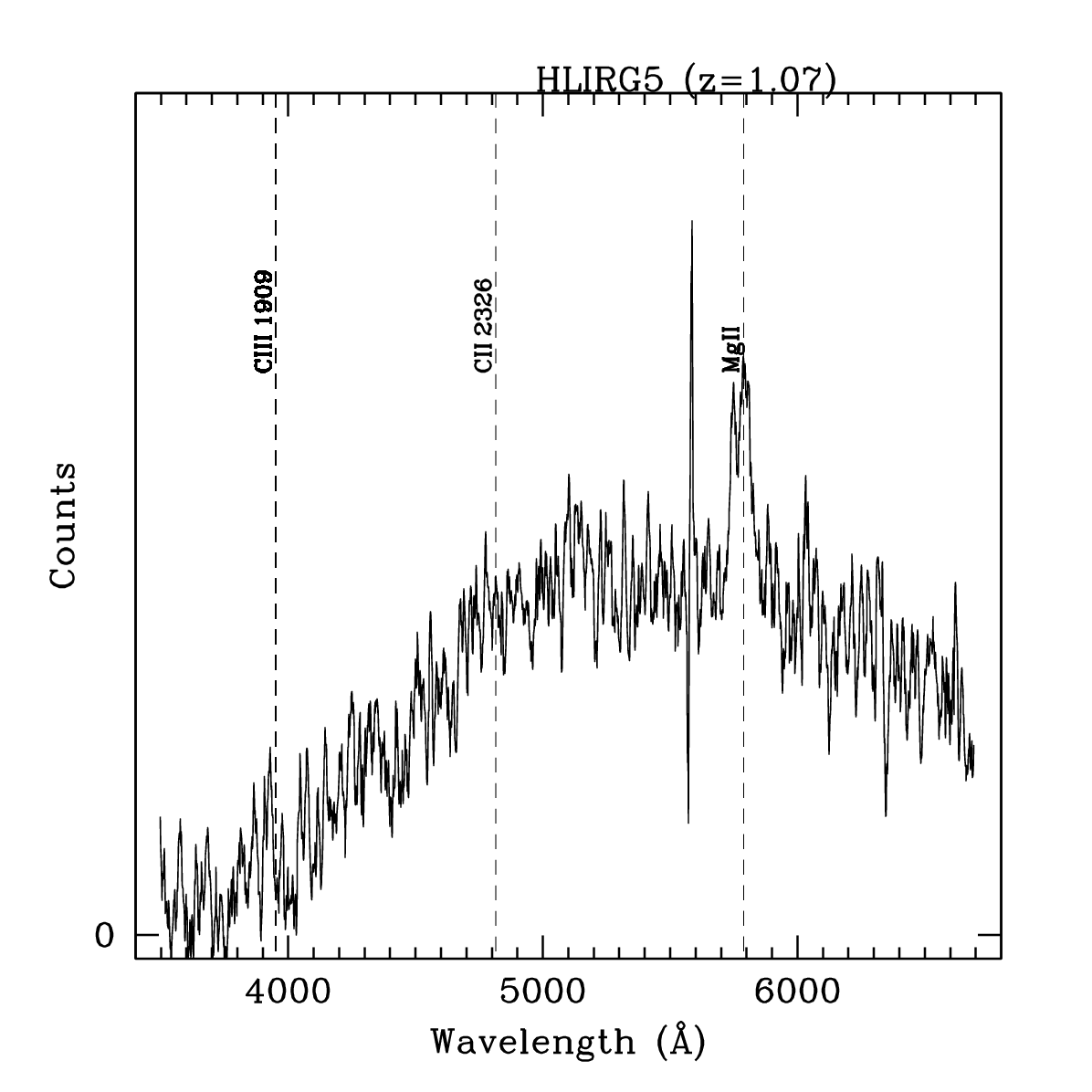}
\includegraphics[width=7. cm]{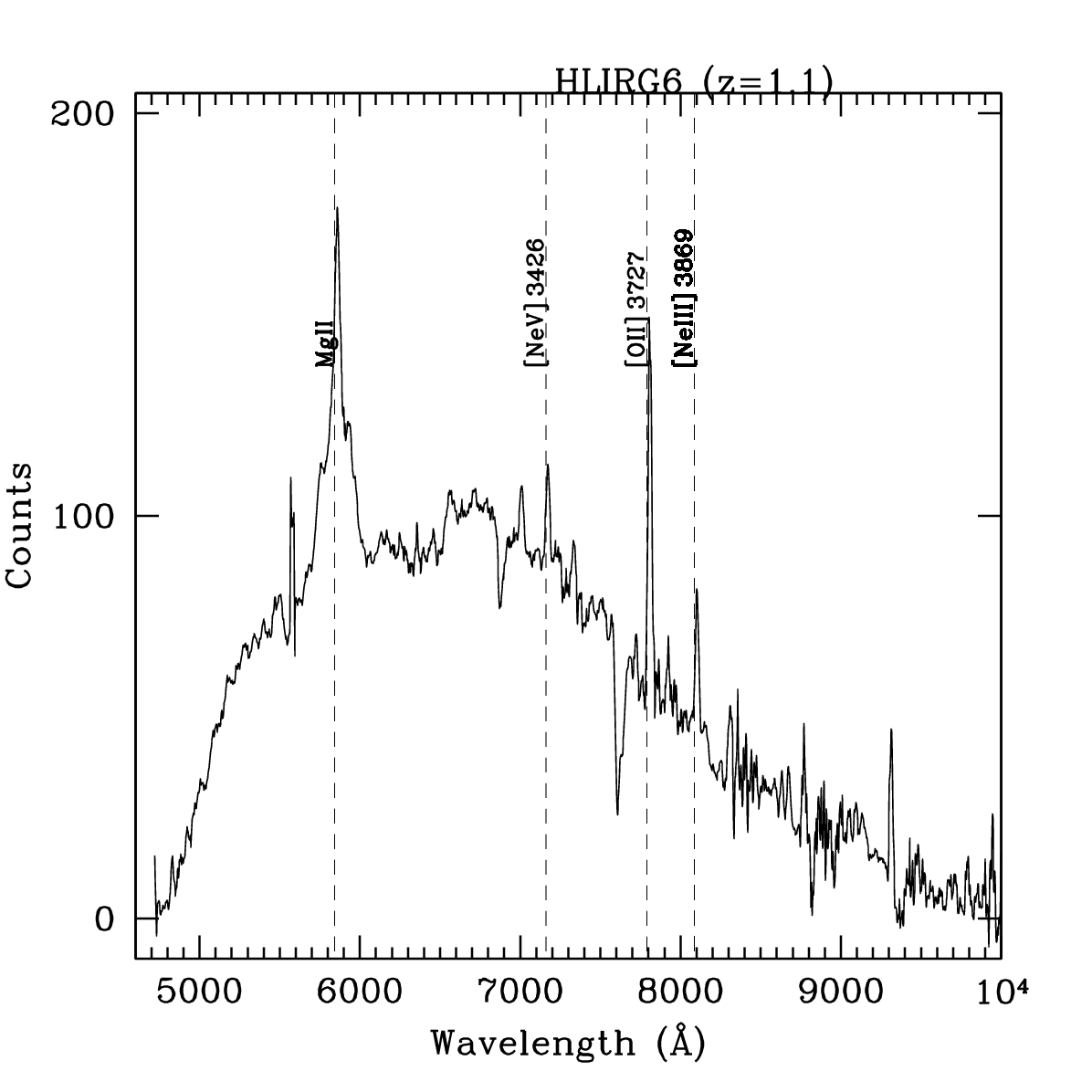}
\caption{Optical spectra of all HLIRGs found in our sample.}
\end{center}
\end{figure*}
\subsection{Data Reduction}
Data reduction of GMOS masks was carried out with the v1.8 GMOS package within IRAF while WIYN observations 
were reduced following the standard procedure for HYDRA. Redshifts were determined by visual inspection of the 
one dimensional spectra, through the identification of emission and absorption features. The presence of strong 
emission and absorption features such as $CIV$, $[CIII]$$\lambda$$1909$, $[MgII]$, $NeV$, $[OII]$$\lambda$$3727$, 
$H$$\beta$, $[OIII]$$\lambda$$4959$, $[OIII]$$ \lambda$$5007$, $H$$\alpha$ , $[NII]$$\lambda$$6583$ directly 
indicated the redshift, epsecially in the cases where at least three of these lines were present. We successfully derived 
197 and 137 redshifts from our GMOS and WIYN runs respectively.  For most of the latter we had both blue and red 
parts, the blue covering the wavelength range 3500--6500~\AA. and the red one 5000--10000~\AA.   
\begin{table*}
\begin{center}
\begin{minipage}{18cm}
\scriptsize
\begin{tabular}{ c*{18}{c}}
\hline
\hline
\multicolumn{1}{c}{ID}&\multicolumn{1}{c}{RA}&\multicolumn{1}{c}{DEC}&\multicolumn{1}{c}{z\footnote{Spectroscopic Redshift}}&\multicolumn{1}{c}{R}&\multicolumn{1}{c}{$S_{24}$\footnote{Observed 24$\rm \mu m$ fluxes in $\rm mJy$}}&\multicolumn{1}{c}{$log(L_{IR})$\footnote{Bolometric Infrared Luminosity($\rm 1-1000 \mu m$) ($\rm H_{0}=72~km~s^{-1}~Mpc^{-1}$,  $\lambda=0.7$)}}&\multicolumn{1}{c}{J1\footnote{Optical SED best fit}}&\multicolumn{1}{c}{J2\footnote{Infrared SED best fit}}&\multicolumn{1}{c}{TC\footnote{Fraction of AGN contribution at 8 $\mu$m}}&\multicolumn{1}{c}{SC\footnote{Fraction of Starburst contribution at 8 $\mu$m}}\\
\hline
\hline
HLIRG-1&242.9055&54.4291&1.392&20.83&5004&13.26&QSO&Torus&1.0&0.0\\
HLIRG-2&242.9449&54.0686&0.690&20.75&1563&13.22&QSO&Torus&0.9&0.1\\
HLIRG-3&242.8916&54.4591&2.860&18.99&1106&13.16&QSO&Torus&0.9&0.1\\
HLIRG-4&242.6961&54.6165&2.700&21.37&0&13.09&QSO&Torus&1.0&0.0\\
HLIRG-5&242.5173&54.2274&1.070&20.00&1906&13.31&QSO&Torus&1.0&0.0\\
HLIRG-6&242.7487&54.3648&1.100&20.44&1123&13.00&QSO&Torus&0.6&0.4\\
\hline
\hline
\end{tabular}
\end {minipage}
\caption{Properties of the 6 Hyper Luminous Infrared Galaxies with available spectra from GMOS/WIYN.}
\end{center}
\end{table*}
\section{Results}
GMOS/WIYN spectroscopic follow-up represents the largest optical spectroscopic sample obtained in ELAIS-N1 to date, with 
a total of 376 objects observed. We were able to successfully determine redshifts for 334 sources, a success rate of 
89$\%$. For the remaining 47 sources a redshift could not be determined due to the bad quality of the spectra or the 
lack of prominent spectral features in the wavelength range covered. Out of the 334 successfully identified objects, 
40 (12$\%$)  were found to have redshift 0 (stars) mostly from the 24$\mu$$\rm{m}$ selection. For the remainder of this work only the 294 objects identified 
as extragalactic sources are used.
A table of our final spectroscopic redshift catalogue is provided in electronic form. We have used a classification system similar to that of  Le Fevre et al. (1995) with zflag =1:  ambiguous solutions due to noisy
spectra; zflag=2: reliable spectroscopic redshifts with at least two lines; zflag=3: highly reliable redshifts with 3 or 
more lines; zflag=9: redshifts based on a single strong line.\\
The redshift distribution of the spectroscopic sample is shown in Figure 1.  The WIYN distribution has a peak at redshift of 0.3--0.4, because many of the sources targeted during this run were the clusters brightest galaxies and also putative cluster members between the two XMM clusters, the regular,Ênorth-eastern and the irregular, south-western cluster (Valtchanov et al. in prep). In addition most of 
the 24  $\mu$$\rm{m}$ sources targeted were also at $z<0.5$.  The lack of objects with redshifts from WIYN 
spectroscopy in the range $0.6<z<0.7$ is most likely due to the limits of the instrument's spectral range.  Objects 
with these redshifts have their OII and H$\alpha$ lines shifted around the upper end of the spectral range. 
A small secondary peak appears for galaxies at redshift z$\sim$1.2. All of these sources have prominent MgII 
lines and are sources which were fitted with dust torii by the photometric redshift code.\\
In the case of the GMOS run there is a peak at z$\sim$0.8 due to the fact that most of the sources selected for this 
sample were selected to have photometric redshifts of $\sim$1 suggesting a very good agreement with the photo-z code used. In addition all of the single strong emission line 
objects which were fitted with an OII line would appear to lie at a redshift range of $0.8<z<1.0$, which even though 
they consist only 5$\%$ of the sample, contribute to the primary peak. A secondary peak which appears at 
low redshifts are sources with strong H$\alpha$ lines selected from the 24  $\mu$$\rm{m}$ SWIRE 
sample. The much lower number of high redshift sources found with GMOS is due to the restricted wavelength 
coverage of the instrument and the lack of usage of a blue grating which resulted in most of the lines shortwards 
[MgII] lying outside the observable range. \\
Figure 1 (right) shows the R-band distribution for the extragalactic sources. Cut-offs are due to the limitations of 
the 3.5 and 8 meter telecopes to obtain spectroscopy of sources up to 21.5 and 23 magnitude respectively, in a 
reasonable amount of time. \\
We have used Rowan-Robinson et al (2008) template fitting method to fit SEDs to all observed sources and 
estimate infrared luminosities. Among these there are 6 hyperluminous infrared galaxies (HLIRGs) with
$L_{IR}$$>$$10^{13} L_{\odot}$. Figure 2 shows the spectra of all HLIRGs observed and Table 2 summarizes their properties. This represents a major new sample of spectroscopically identified HLIRGs.  There are just 57 hyperluminous infrared galaxies with spectroscopic redshifts in the Imperial IRAS-FSC Redshift  Catalogue (Wang $\&$ Rowan-Robinson 2008).
\subsection{HLIRG}
Here we discuss the properties of the 6 HLIRGs in  our sample.\\
$\bf{HLIRG1:}$ This source, which is one of the brightests of our sample, was observed with WIYN because of its bright R-band magnitude (R=20.83). As it can be seen by its spectra in Figure 2 it exhibits a broad [MgII] line at a redshift of 1.392 confirming the prediction of both optical and IR SED fitting that it is dominated by an AGN. Our observations show a marginal detection of [OII] and FeII lines. It has detections in all IRAC bands with very strong 24 $\mu$$\rm{m}$ emission. The fact that this object has very reliable redshift estimation in addition to the very strong 24 $\mu$$\rm{m}$ detection implies that  the bolometric infrared luminosity estimation is very reliable. \\
$\bf{HLIRG2:}$ This is also one of the bright (R=20.75) sources observed with WIYN. All the Balmer ($H\gamma$, $H\beta$) lines detected are quite broad with a possible detection of [MgII] at the edge of the wavelength coverage. Very strong [OIII] doublet, [OII] and NeIII are clearly visible indicating also the presence of a starburst event present at the same time. This source lies at intermediate redshift (0.69) and is also a very strong 24 $\mu$$\rm{m}$ emitter. Both optical and IR SED fitting imply the presence of a strong AGN , as confirmed by the broad lines, but with significant starburst contribution as also seen from its spectrum. Precise redshift and 24 $\mu$$\rm{m}$ measurements make it an additional very strong candidate for HLIRG status. \\ 
$\bf{HLIRG3:}$ This is the brightest example (R=18.99) in the sample of HLIRG at very high redshift (z=2.86). We have detected at least 2 broad lines CIII and CIV, the possible detections of two more SiIV and HeII and the detection of Ly$\alpha$ in the blue. The clear presence of these lines makes it almost certain the presence of a very strong AGN which is also identified by optical and IR SED fitting. The latter though implies the presence of rather strong starburst at this early stage of evolution.\\
$\bf{HLIRG4:}$ This source exhibits three broad lines CII, CIII and CIV. The clear detection of these lines indicates the presence of a QSO in agreement with SED fiiting. It is the only source that lacks 24$\mu$m detection.\\
$\bf{HLIRG5:}$This object exhibits a very broad MgII line and marginal detections of CII and CIII. The clear presence of the MgII broad line in the slit makes the redshift estimation almost certain. It is a very strong 24$\mu$m emitter with detections in all IRAC and MIPS bands, except 160$\mu$m. Its SED shows an AGN powered source with no evidence of on-going star-formation.\\ 
$\bf{HLIRG-6:}$ This object exhibits four strong lines, a broad [MgII] line accompanied by the presence of NeV, [OII] and NeIII. Spectroscopy and photometry agree to the fact that this is a composite object with clear presence of both an AGN and a star-formation component.
\section{Comparison with photometric redshifts}
We have compared the spectroscopic redshifts obtained with GMOS/WIYN with photometric redshifts
calculated using the latest version of the ImpZ code (Rowan-Robinson et al. 2008). Reliability and accuracy of the 
photometric redshifts are measured via the fractional error $\Delta z/(1+z)$ for each source, examining the mean 
error $\overline{\Delta} z/(1+z)$, the $rms$ scatter $\sigma_z$ and the rate of `catastrophic' outliers $\eta$, defined 
as the fraction of the full sample that has $|\Delta z/(1+z)|>0.25$.\\
Figure 3 shows a comparison of $log_{10}(1+z_{phot})$ versus $log_{10}(1+z_{spec})$ for the entire sample of 
SWIRE ELAIS-N1 sources with available spectroscopy from GMOS/WIYN, where red diamonds are galaxies and 
blue triangles are QSOs, as reported by the photometric optical template fitting.  For galaxies
with confidence class zflag=1, where there are ambiguous redshift solutions due to noisy
spectra, we tried to use the information 
from the optical photometry to select between the aliases, where possible. ImpZ code found solutions for all sources. 
In the case of zflag=2 or 3 objects, the photometric code was successfull at returning an accurate z$_{phot}$ 
with $\sigma_{tot}$=0.09, $\overline{\Delta} z/(1+z)=-0.008$ and only 4.4$\%$ (10 sources) catastrophic outliers. 
One of the key results here is the excellent performance of the photometric code for both sources identified as QSOs 
from the optical fitting and for high redshift sources with good quality of spectra. From the 19 sources with spectroscopic 
redshift $z>1.5$  there are only two catastrophic outliers. In addition from the 18 sources which were fitted with a QSO 
optical template the number of outliers was two. Taking into consideration the problems that photometric codes face 
with high redshift sources and QSOs and bearing in mind that the main power of photometric redshifts is their 
application in both these populations and especially in high redshift sources then this success is encouraging. In terms of outliers, possible reasons for these are either photometric redshift aliasing or poorer 
optical photometry for the case of the GMOS fainter sample.  The photometric code outputs the $\chi^2$ distribution 
as a function of redshift and from this we can assess whether photometric redshift aliases are likely. This suggested 
that only for one of the 10 sources with $|log_{10}((1+z{phot})/(1+z_{spec}))
> 0.10$ is there an obvious photometric alias.  The main cause for the remaining outliers in this sample is the 
worsening photometric accuracy at the fainter optical magnitudes sampled here. \\
\begin{figure*}
\begin{center}
\includegraphics[width=8. cm]{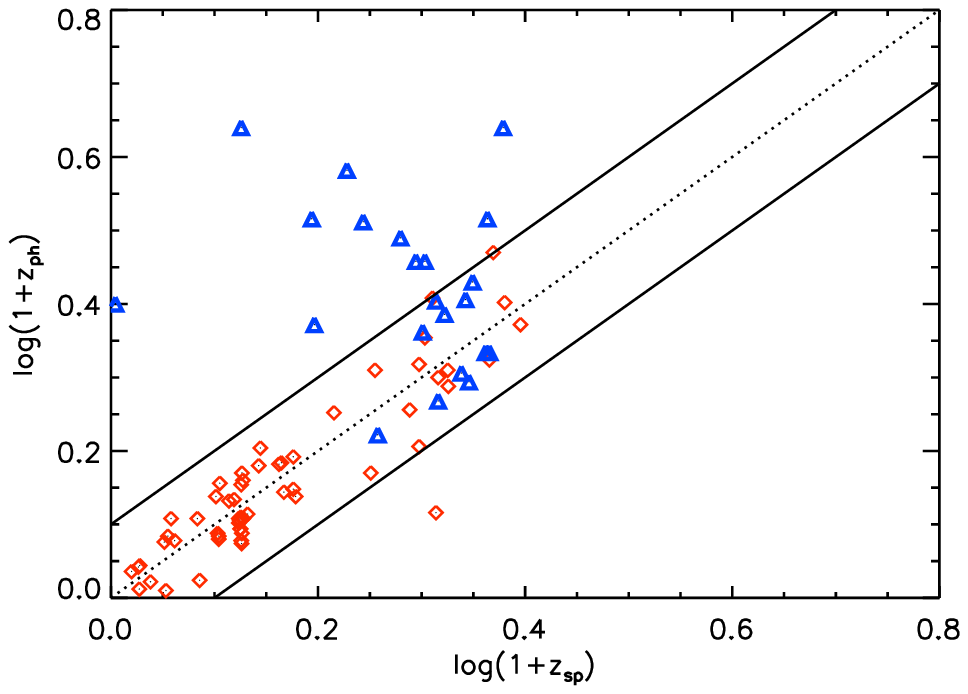}
\includegraphics[width=8. cm]{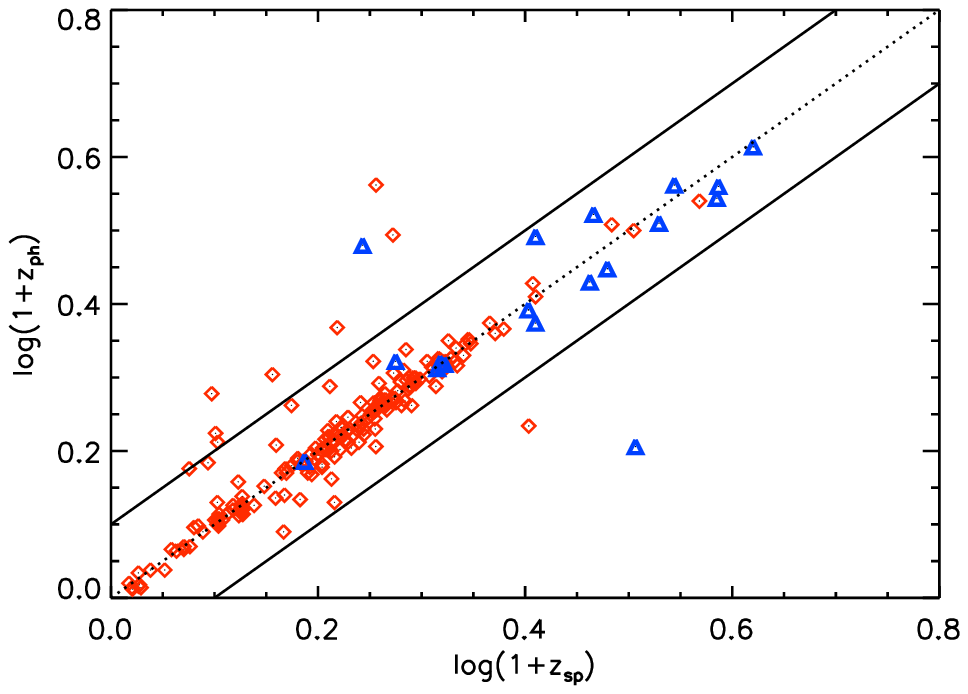}
\caption{$\bf{Left:}$Photometric versus spectroscopic redshift for all sources belonging either to zflag=1 or 9 (ie noisy spectra or single-line redshifts). The straight lines represent a $\pm$ 0.1 dex accuracy in log(1+z). Blue triangles are sources 
fitted with a QSO template and red diamonds are sources fitted with a galaxy template (Rowan-Robinson et al. 2008). $\bf{Right:}$ Photometric versus
 spectroscopic redshift for all sources belonging to the secure spectroscopic  confidence classes zflag=2 and 3. Colours and 
symbols are the same as left-hand figure.}
\end{center}
\end{figure*}
\begin{figure*}
\begin{center}
\includegraphics[width=7.2cm]{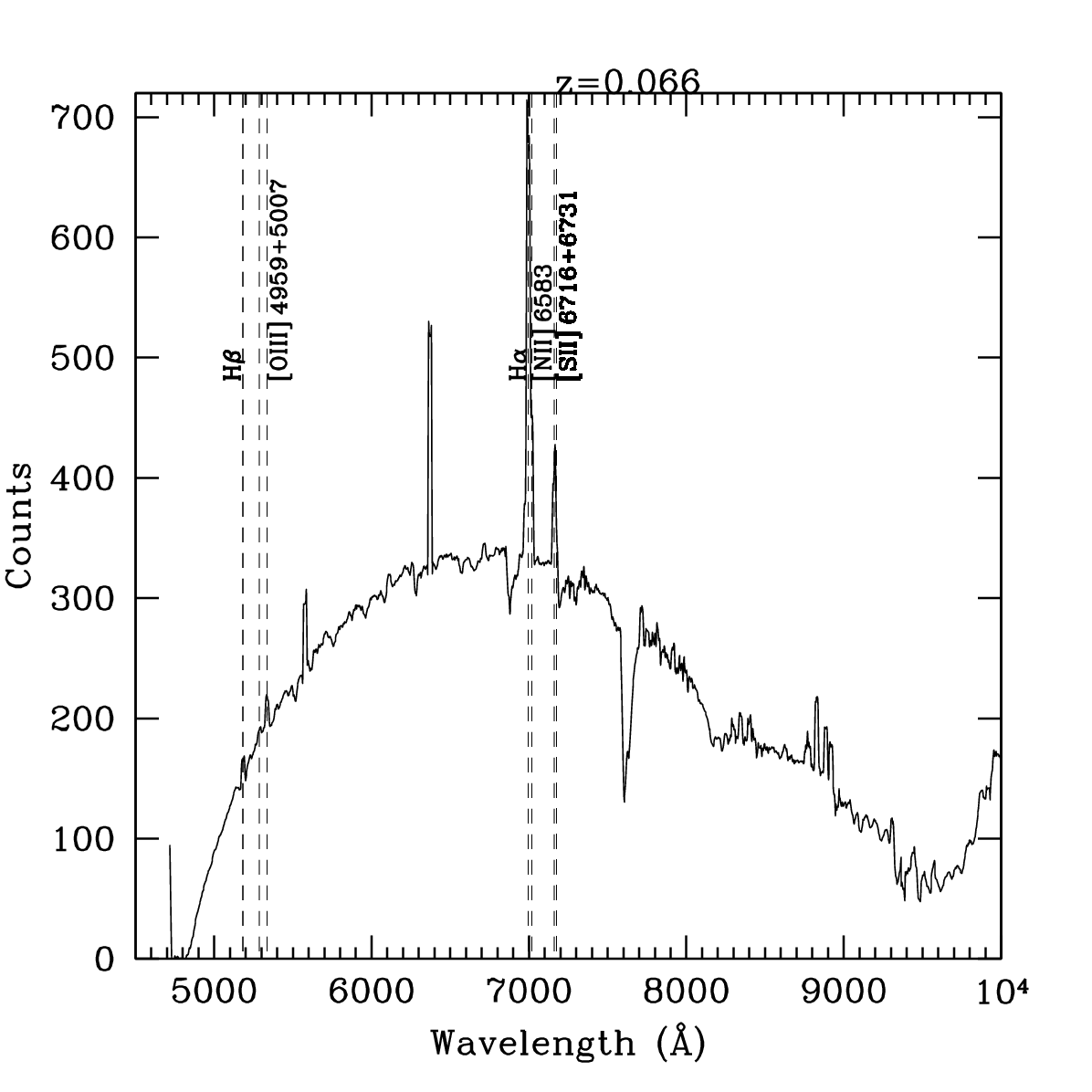}
\includegraphics[width=7.2cm]{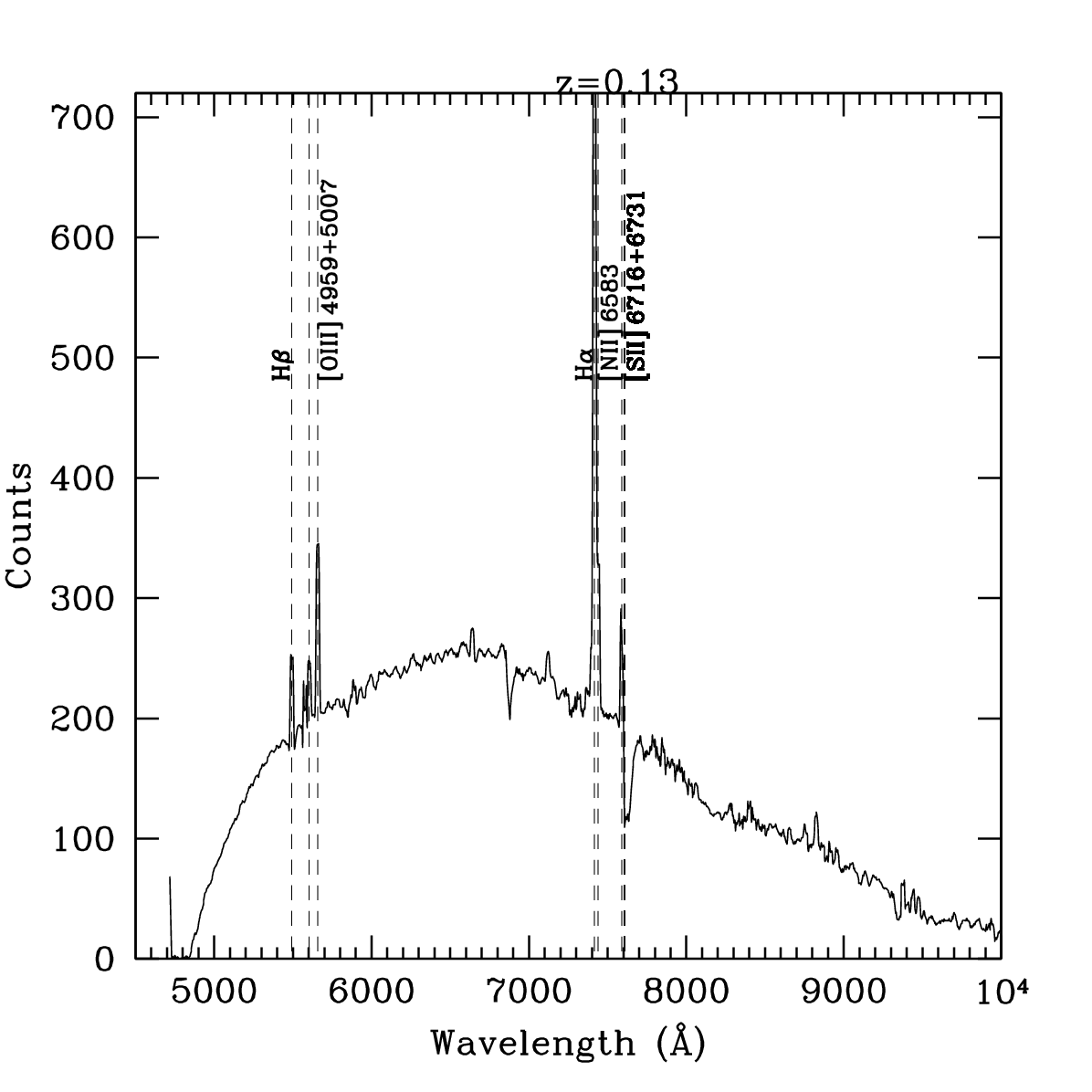}
\includegraphics[width=7.2cm]{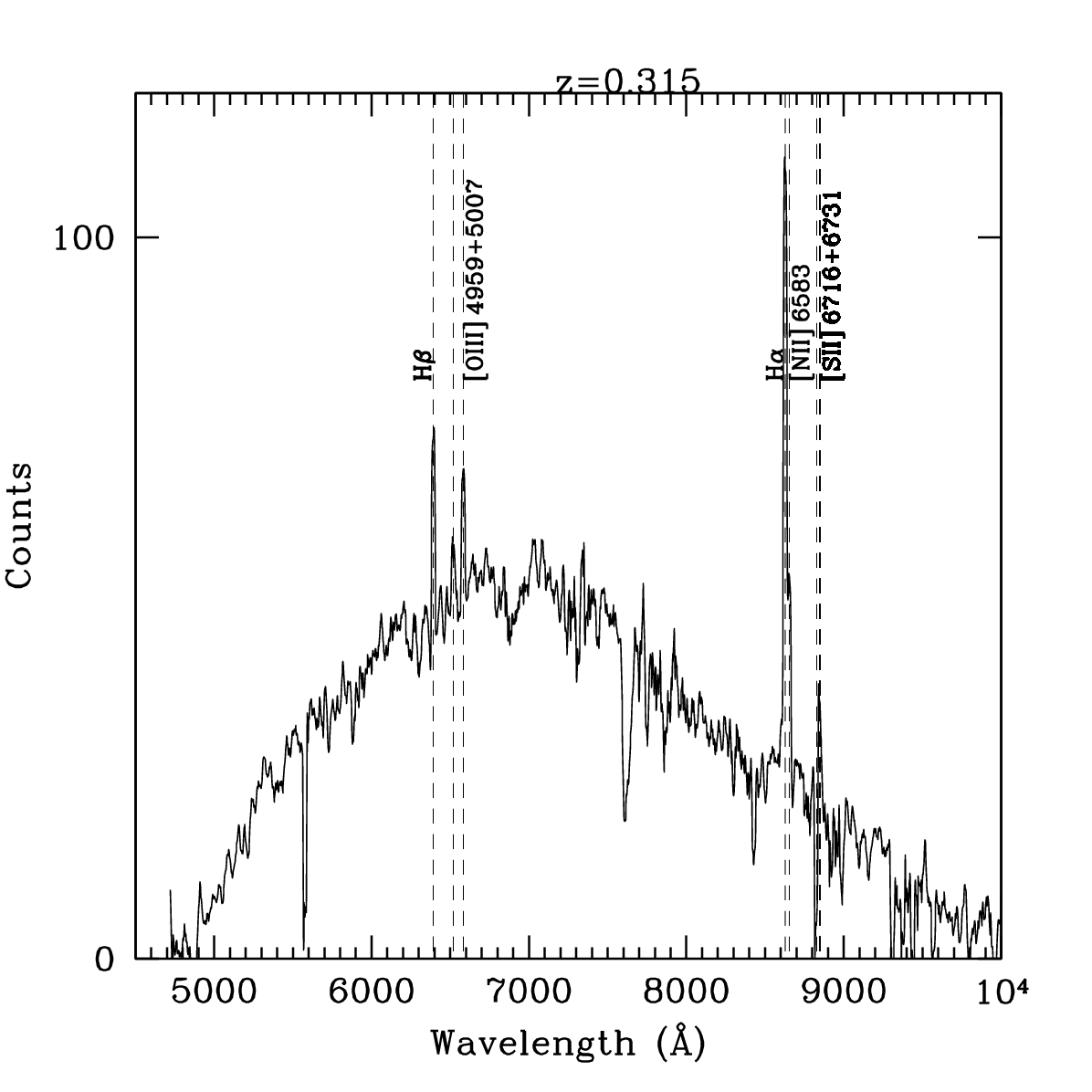}
\includegraphics[width=7.2cm]{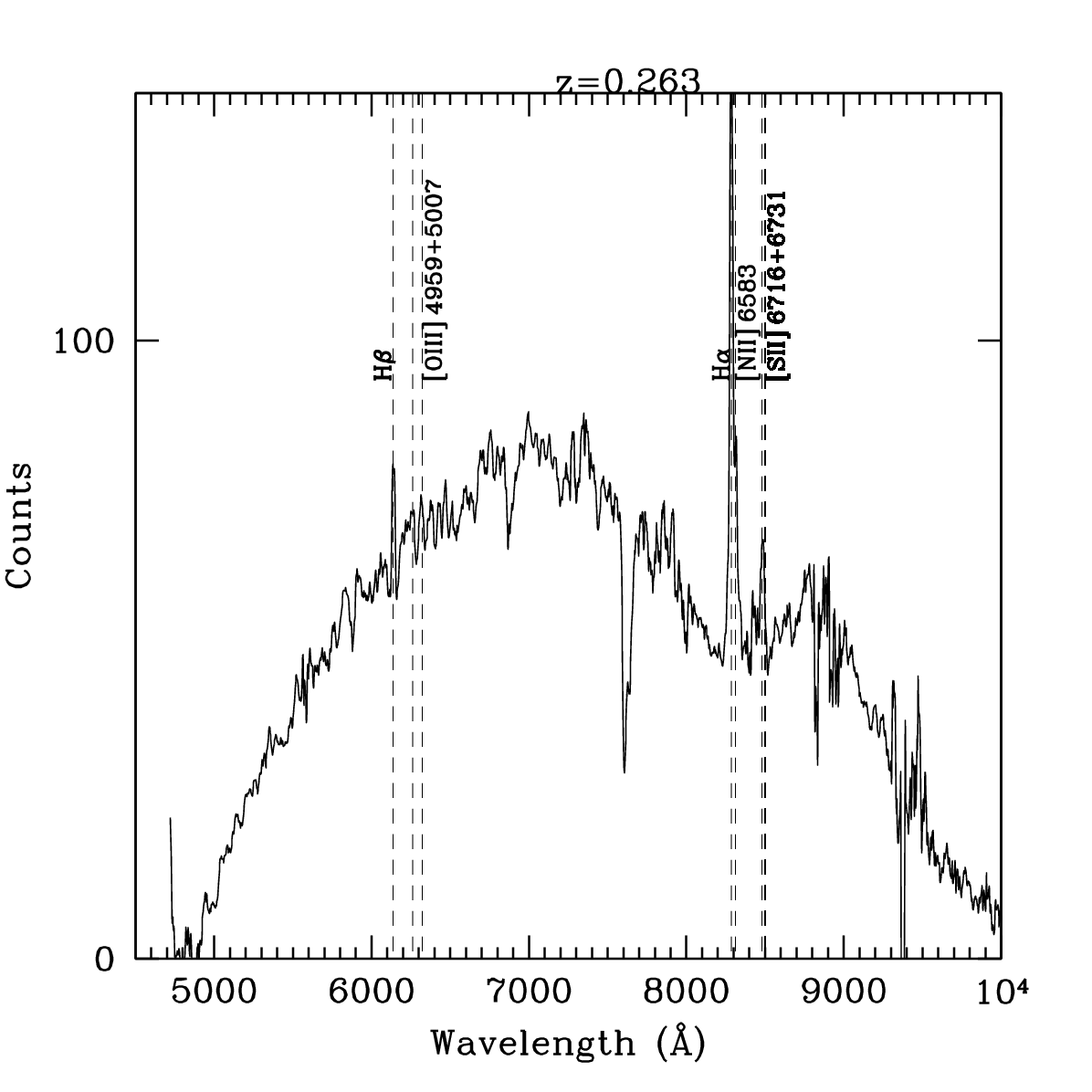}
\includegraphics[width=7.2cm]{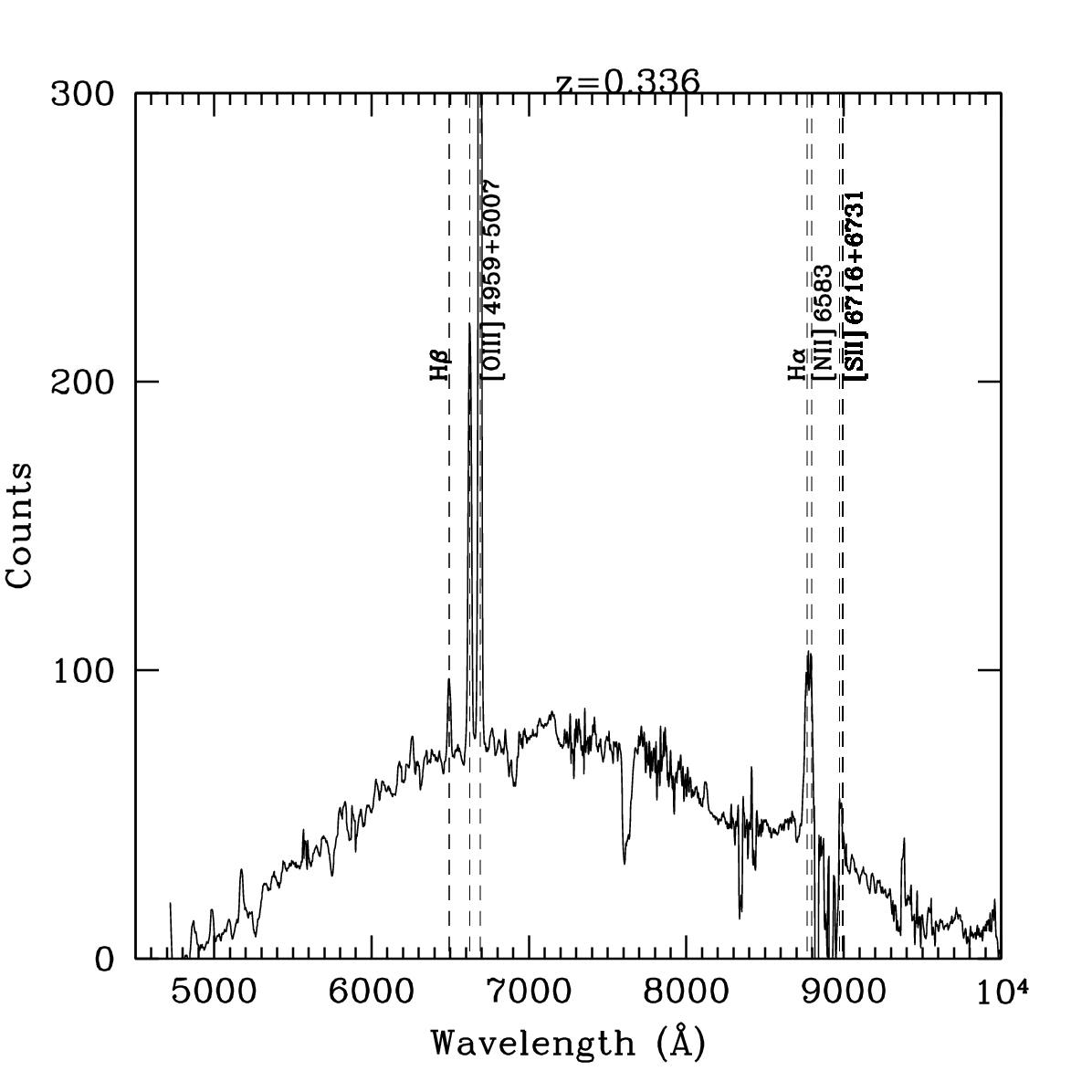}
\includegraphics[width=7.2cm]{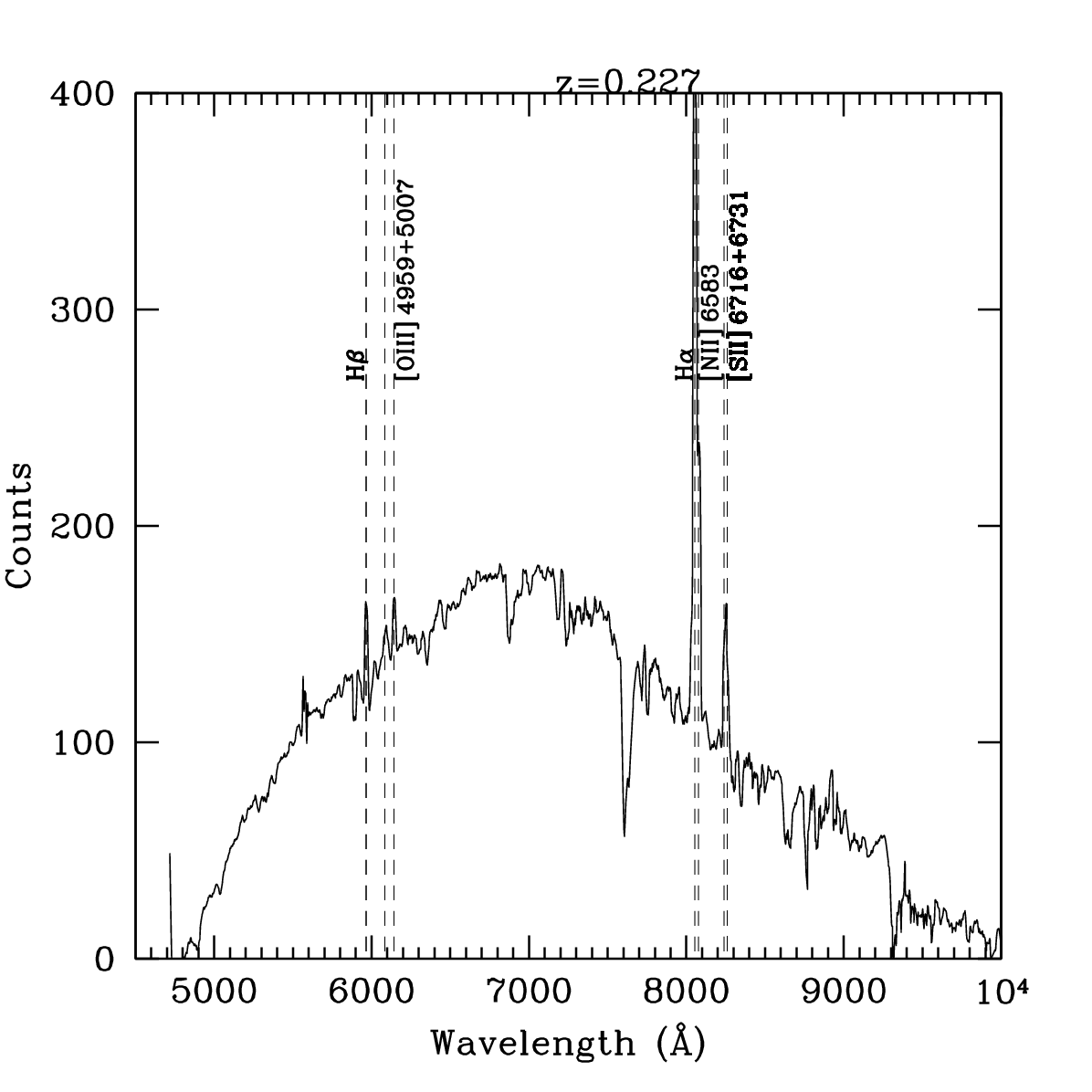}
\caption{Examples of spectra with available $[SII]$, $H$$\alpha$, $[OIII]$, $H$$\beta$, $[NII]$ lines, used to estimate line ratios.}
\end{center}
\end{figure*}
\begin{figure*}
\begin{center}
\includegraphics[width=8.8 cm]{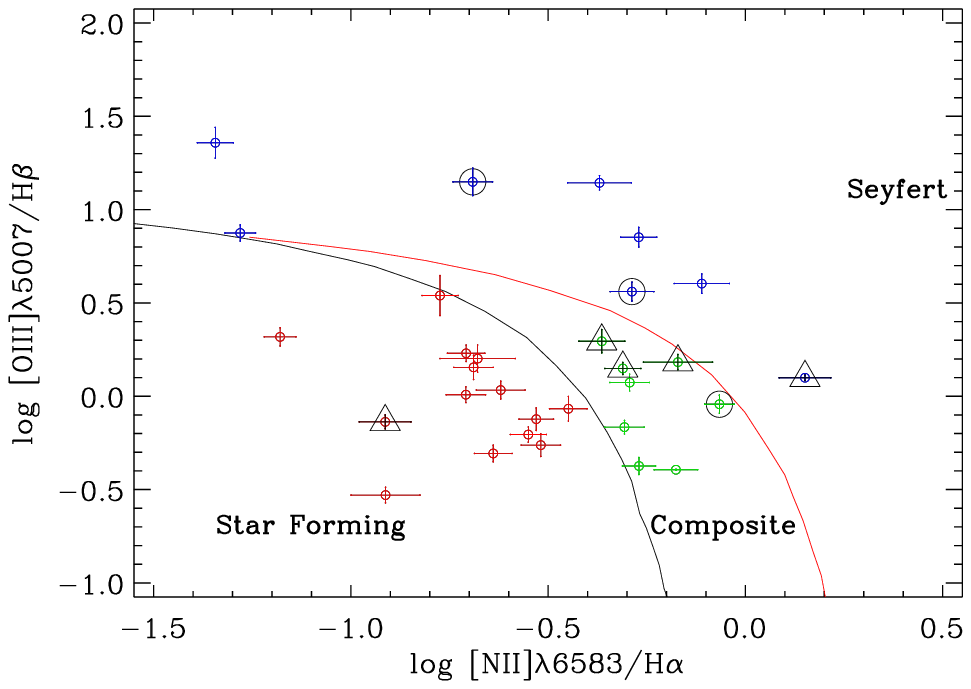}
\includegraphics[width=8.7 cm]{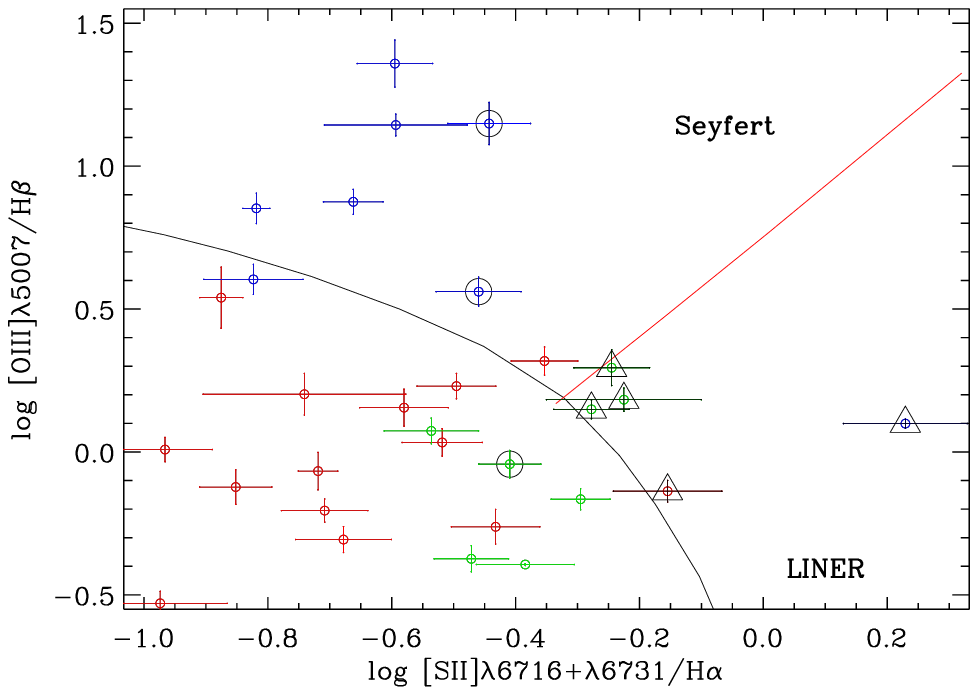}
\caption{$\bf{Left:}$ The $[NII]$/$H$$\alpha$ versus $[OIII]$/$H$$\beta$  diagnostic diagram for all 30 sources with available lines. The black line is the pure star formation line (Kauffmann et al. 2003; Brinchmann et al. 2004).  Red line is the extreme starburst line  (Kewley et al. 2001). $\bf{Right:}$ The $[SII]$/$H$$\alpha$ 
versus $[OIII]$/$H$$\beta$  diagnostic diagram for all 30 sources with available lines. The red line is the Seyfert/LINER line (Kewley et al. 2006) and the black line is the AGN/starburst line (Baldwin et al. 1981). $\bf{Both:}$ Open black circles are sources with X-ray detections from the ELAIS-N1 Deep X-ray survey (Manners et al. 2003). Red, blue and green colors represent sources identified as star-forming, Seyfert and composite respectively, based on the $[NII]$/$H$$\alpha$ versus $[OIII]$/$H$$\beta$  diagnostic. Open black triangles are sources identified as LINERs based on the $[SII]$/$H$$\alpha$ versus $[OIII]$/$H$$\beta$  diagnostic.}
\end{center}
\end{figure*}
For the case of sources flagged as having ambiguous spectroscopic redshift estimations (zflag=1 and 9, Fig 3 left)
the agreement is broadly good, but it is noticeable that both in terms of the rms accuracy and in the number of serious 
outliers the results look worse than those presented for zflag=2 and 3, suggesting that some of the increased
scatter in Figure 3 (Left) compared to Figure 3(right) is due to incorrect spectroscopic redshifts.  The percentage of catastrophic outliers is 
15.4$\%$, of which 92$\%$ are sources fitted with an optical QSO template. For sources with $z_{photo}<0.6$ we 
used the information from the optical photometry to select between aliases with relative success. The parameter 
space between $0.7<z_{photo}<1.5$ is occupied mainly by the sources given a confidence class of 9. These are 
the very strong single emission line objects fitted with an OII line. In the literature people tend to assign these objects 
either to OII or H$\alpha$ lines (e.g. Le Fevre et al. 1995). In our case assignment to OII lines provides a good 
agreement with the photoz values with no catastrophic outliers. Significant disagreement occurs at  $z_{phot}>1.5$ 
where most of our outliers occur.  In this redshift range most of the prominent spectral features would be 
redshifted outside the wavelength coverage of both instruments with the set-up used here. 
\section{Emission Line Diagnostics}
A suite of three emission line diagnostic diagrams has been used to classify the dominant energy source in narrow 
emission line galaxies (Baldwin et al. 1981). These diagrams are based on the three optical line ratios: 
$[OIII]$/$H$$\beta$,  $[NII]$/$H$$\alpha$ and $[SII]$/$H$$\alpha$.  Although other ratios, 
such as $[OII]/[OIII]$, offer cleaner separation between planetaries and $H_{II}$ regions (Baldwin et al. 1981),  
combinations such as $[NII]$/$H$$\alpha$ versus $[OIII]$/$H$$\beta$ and  $[SII]$/$H$$\alpha$ versus 
$[OIII]$/$H$$\beta$  not only are very accurate in distinguishing the dominant component of narrow line galaxies 
but are also very insensitive to reddening and do not require the samples to be flux calibrated. This is due to the 
fact that  the lines in these pairs are chosen to be close in wavelength and as a result the line ratios are accurately 
determined even for non flux calibrated spectra.  Kewley et al. (2001 $\&$ 2006) used SDSS spectra to improve Baldwin et al  (1981) diagnostic diagrams. They showed that the LINER/Seyfert divide in the [OIII]/Hbeta vs [NII]/Halpha plot is more indicative of composite star forming/AGN activity than a true LINER nature, and that the [SII]/Halpha  axis is better at distinguishing the LINERS. \\ 
We have identified a total of 30 sources with very secure detections of all 5 lines, $[OIII]$, $H$$\beta$,  $[NII]$, 
$H$$\alpha$, $[SII]$.  The flux in each emission line is measured using IRAF's SPLOT routine, by interactively 
fitting a Gaussian function to each emission feature line profile and then integrating.  Figure 4 shows examples of the spectra of these objects.\\
Figure 5 shows the $[NII]$/$H$$\alpha$ versus $[OIII]$/$H$$\beta$ and $[SII]$/$H$$\alpha$ versus 
$[OIII]$/$H$$\beta$ diagnostic diagram for our sample of 30 sources. The black line represents the most conservative AGN rejection criterion and they are the sources with very low contribution to $H$$\alpha$ from AGN (Brinchmann et al. 2004). This line is similar to Kauffmann et al. (2003) pure starburst line which has included this line to distinguish between pure star-forming sources and composite objects whose spectra contain significant contributions from both AGN and star formation. Everything below this line are classed as HII region like galaxies. The red line in the $[NII]$/$H$$\alpha$ versus $[OIII]$/$H$$\beta$ diagnostic is the line developed by Kewley et al (2001). This line uses a combination of stellar population synthesis models and detailed self consistent photoionization models to create a theoretical maximum starburst line. Everything that lies above this line should be AGN dominated. All sources that lie between these two lines are classed as composites and their spectra can be either due to a combination of star formation and a Seyfert nucleus or due to a combination of star formation and LINER emission. Figure 5 (right) shows the $[SII]$/$H$$\alpha$ versus $[OIII]$/$H$$\beta$ diagnostic diagram for the same sample as before.  The black line separates between AGN and star forming galaxies while the red line provides an empirical division  between LINER and Seyfert sources (Kewley et al. 2006).\\
Based on the $[NII]$/$H$$\alpha$ versus $[OIII]$/$H$$\beta$ diagram from the 30 sources in our sample we have found 14 pure starforming sources,  8 Seyferts  and 8 composite objects.  According to the $[SII]$/$H$$\alpha$ versus $[OIII]$/$H$$\beta$ diagnostic, 1 Seyfert, 3 composites and 1 star-forming sources appear to be LINERs.
\section{Colour-Colour Diagrams and Spectral Energy Distributions}
In this section we utilize SWIRE infrared data and GMOS/WIYN spectroscopic data to reproduce  the Lacy et al (2004) IRAC color-color plot. For the Lacy et al (2004) diagnostic we require sources to have detections in all four 
IRAC bands, 3.6, 4.5 , 5.8 and 8$\mu$$\rm{m}$. Figure 6 shows the Lacy et al (2004) diagnostic diagram for the sample of the 30 narrow line galaxies and the 35 broad line QSOs found in the GMOS/WIYN spectroscopic catalogue. Sources that lie within the black line are those expected to be AGN dominated from the IRAC colours (Lacy et al. 2004). \\
All broad line objects lie within the AGN wedge as predicted by Lacy et al (2004) and more than half of them (19/35) have been detected 
at X-ray wavelengths. In addition 33/35 broad-line QSOs are fitted with a QSO template at optical wavelengths. 
The remaining two were
fitted with a starburst template and are the only ones in this broad line sample which have broad Balmer lines 
and evidence of star formation. Indeed the SEDs of these two objects are composite and show both an AGN and 
a starburst component with the starburst dominating at 8 $\mu$$\rm{m}$. \\
Turning to the population of narrow emission line galaxies, the majority of the star-forming sources lie on the top left of the IRAC color-color parameter space and outside the AGN wedge in agreement with previous results which utilised 24$\mu$m selected samples (Donley et al. 2008). In the case of the narrow-line AGN, 5 lie within the AGN wedge, 2 occupy the same parameter space as the star-forming sources and only one lies in the Type-2 parameter space, as this was defined by Donley et al (2008) for low-z narrow line AGN. Composite objects and LINERs do not occupy a specific part of the plot. Figure 7 shows the IRAC color-color diagnostic from Stern et al. (2005). Similarly to Lacy et al (2004), the diagnostic is effective in constraining the broad-line objects but not the Type-II Seyferts.\\
Lacy et al. (2007) studied a similar sample of low-z (z$\le$0.3) 24 $\mu$m sources with optical spectra and found some narrow-line AGN in the top left region of the plot. Based on their  strong [OII] lines and composite nature, Lacy et al (2007) concluded that the 8$\mu$m flux was being enhanced by PAH emission from star-formation of high enough EW to affect their IRAC colors which is consistent with what we find in the case of only one of our narrow line AGN that lies outside the wedge. The remaining 2 show evidence of a quiescent galaxy template that dominates their infrared emission. \\
Figure 8 shows the SEDs for all 13 narrow-line objects that lie outside the wedge and occupy the top left region of the IRAC color-color plot and table 3 summarizes their properties. All of them are fitted with the standard infrared templates of Rowan-Robinson et al 
(2004, 2005, 2008).  The templates used are (1) infrared 'cirrus', optically thin emission 
from interstellar dust illuminated by the galaxy's radiation field, (2) an 
M82-like starburst, (3) a more extreme, Arp220-like, starburst, (4) an AGN 
dust torus. All 13 objects  are either fitted with an infrared starburst template or a quiescent galaxy 
('cirrus') template. None have detectable AGN dust tori at infrared wavelengths or X-ray detections. 
For the two Type-2 Seyferts, non-detection in X-rays suggests that these are either very low-luminosity AGN or that the central engine is 
obscured by a high column of dust. The fact that these narrow line AGN are low-redshift 
objects, with infrared colours typical of quiescent or starburst galaxies and
with no evidence of dust tori, and that there is no evidence for a QSO contribution in the optical continuum, suggests 
that these are probably low-luminosity 
AGN completely missed by both X-rays and infrared colour diagnostics.  The only evidence 
for  an AGN comes from the presence of narrow emission lines with the appropriate line-ratios.
\begin{figure}
\begin{center}
\includegraphics[width=8.8 cm]{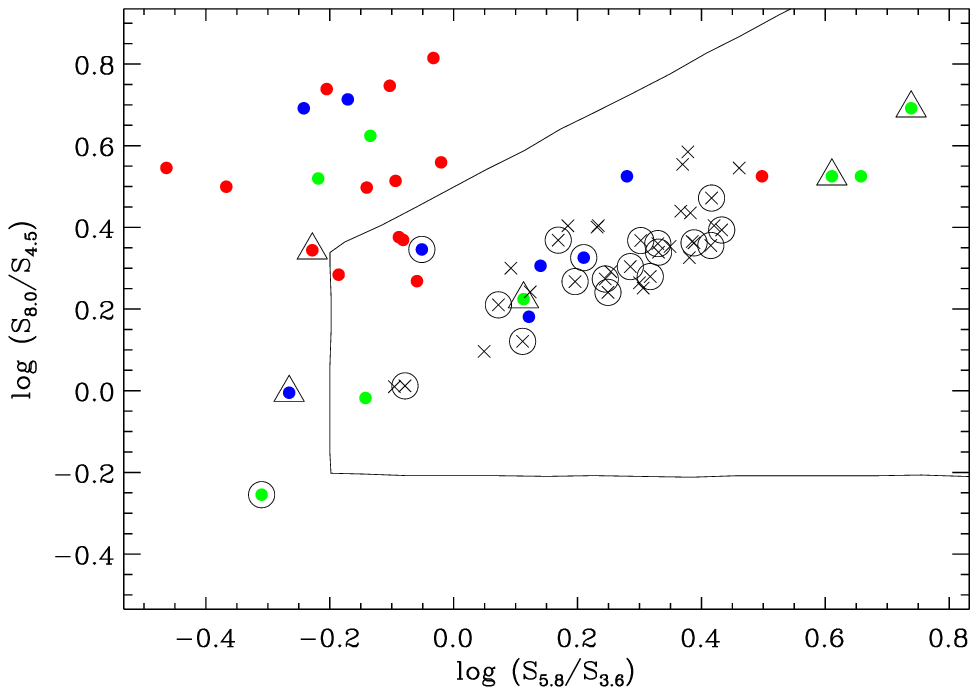}
\caption{IRAC color-color plot (Lacy et al. 2004) of all spectroscopically classified GMOS/WIYN sources with detections in all IRAC bands, namely the 30 narrow line emission sources and the 35 broad line QSOs. Open black circles are sources with X-ray detections (Manners 
et al. 2003). Black crosses are broad line QSOs. Red, blue and green colors are sources identified as star-forming, Seyfert and composite respectively, based on the $[NII]$/$H$$\alpha$ versus $[OIII]$/$H$$\beta$  diagnostic. Open black triangles are sources identified as LINERs based on the $[SII]$/$H$$\alpha$ versus $[OIII]$/$H$$\beta$  diagnostic.}
\end{center} 
\end{figure}

\begin{figure}
\begin{center}
\includegraphics[width=8.8 cm]{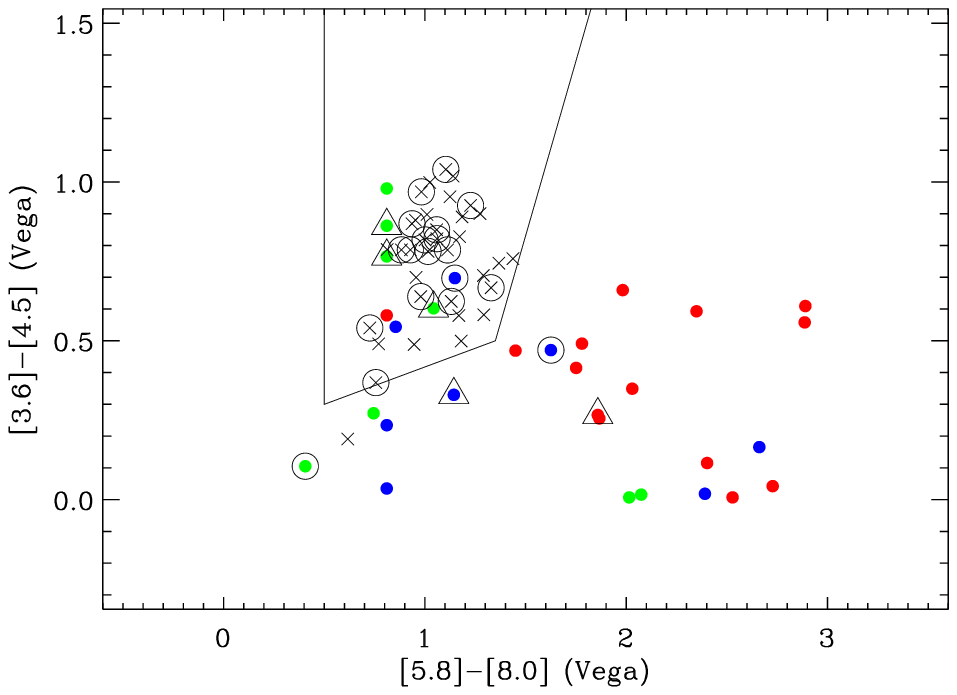}
\caption{IRAC color-color plot (Stern et al. 2004) of all spectroscopically classified GMOS/WIYN sources with detections in all IRAC bands, namely the 30 narrow line emission sources and the 35 broad line QSOs. Open black circles are sources with X-ray detections (Manners 
et al. 2003). Black crosses are broad line QSOs. Red, blue and green colors are sources identified as star-forming, Seyfert and composite respectively, based on the $[NII]$/$H$$\alpha$ versus $[OIII]$/$H$$\beta$  diagnostic. Open black triangles are sources identified as LINERs based on the $[SII]$/$H$$\alpha$ versus $[OIII]$/$H$$\beta$  diagnostic.}
\end{center} 
\end{figure}

\begin{figure}
\begin{center}
\includegraphics[width=8. cm]{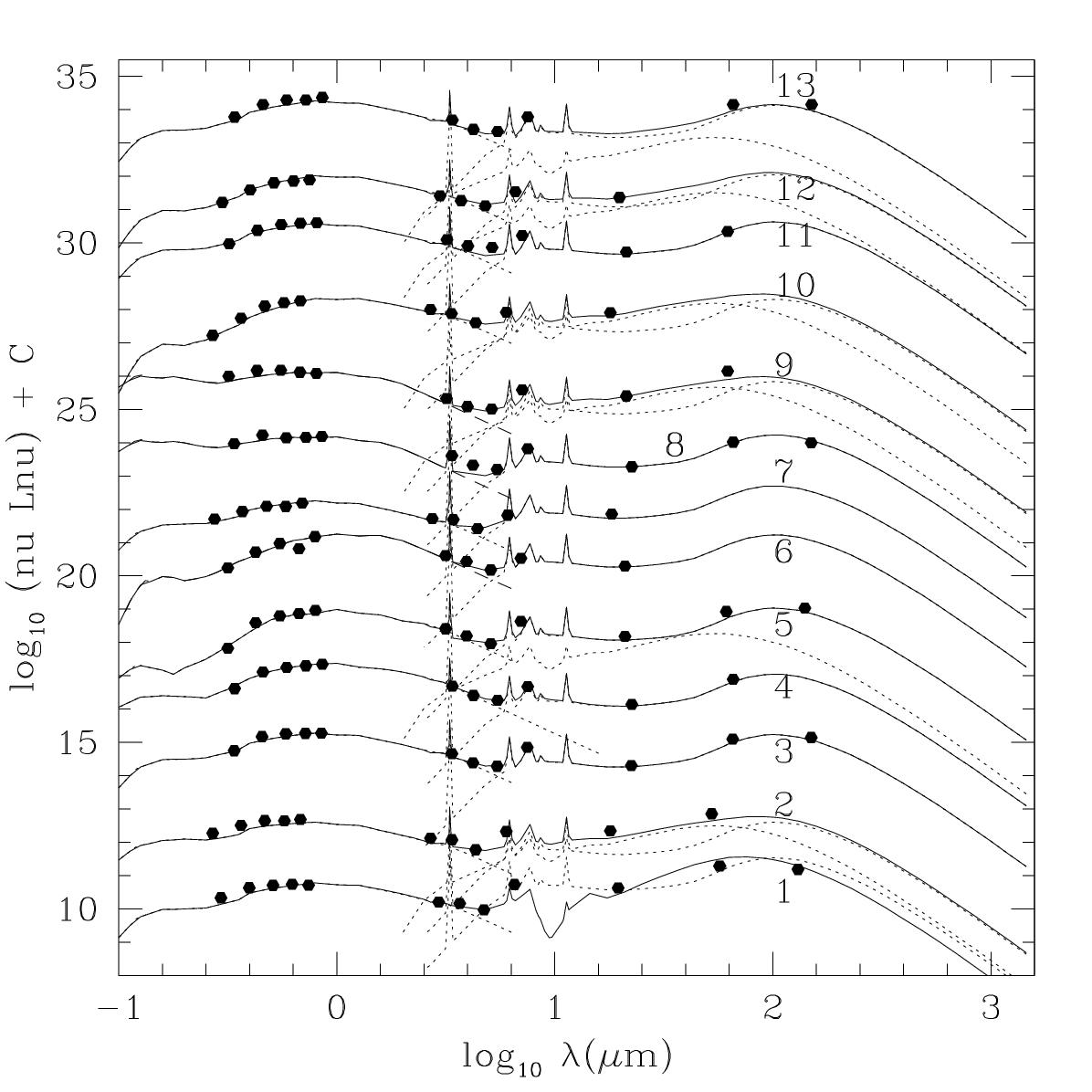}
\caption{ SEDs for the 13 narrow emission-line galaxies  that occupy the star-formation region of the Lacy et al (2004) diagnostic.  7 are fitted with a pure
'cirrus' template (objects 3-8, 11), 5 require cirrus + an M82 starburst template (objects 2, 5, 9, 10, 12, 13) and
one (obect 1) is fitted with an Arp 220 starburst template.  Parameters for all 13 galaxies are given in Table 2.}
\end{center} 
\end{figure}
\section{Summary}
We present the largest spectroscopic follow-up performed in SWIRE ELAIS-N1 using the Gemini Multi-Object 
Spectrograph on Gemini-North and the Hydra multi-object spectrometer on the WIYN telescope. The values of 
spectroscopic redshifts of the later have been compared with the estimated values from our photometric redshift 
code with very good agreement between the two for both galaxies and quasars, even for sources with very faint 
optical magnitudes. The 
scatter between spectroscopic and photometric redshifts was worse for
objects with poorer quality spectroscopic data, suggesting that at least part of this increased scatter is due
to incorrect spectroscopic redshifts.\\
We identified 6 HLIRGs with spectroscopic redshifts, all of which had QSO spectra in the optical, and mid-infrared SEDs
dominated by dust tori.\\
For 30 galaxies with 5 diagnostic lines present ($[OIII]$, $H$$\beta$, $[NII]$, $H$$\alpha$ and $[SII]$) we have classified the galaxies into star-forming, Seyfert, composite and LINER. Broad-line AGN all show
dust torus components in the infrared and this explains why they lie within the infrared colour-space defined
by Lacy et al (2004).  
However  narrow line AGN do not all lie in the colour-space proposed by Lacy et al. (2004) and Donley et al. (2008), in agreement with Barmby et al. (2006) who showed that mid-IR diagnostics cannot identify optical Type-II AGN. Infrared SED fitting 
suggests that the Type-II sources that do not lie in the proposed AGN parameter spaces are all starburst dominated or quiescent galaxies. Any dust torus present must be very weak.
\section{ACKNOWLEDGMENTS}
The authors wish to thank the SWIRE team for their constructive comments on the original GMOS and WIYN ELAIS-N1 proposals. Also the authors would like to thank Mark Lacy and the anonymous referee for their useful comments. This work has been supported by funding from the Peren grant (MT) and the STFC Rolling grant (MT).
\begin{table*}
\caption{Parameters for narrow emission-line AGN in our spectroscopic survey, see Fig 8. The parameters have been estimated using the SED template fitting method from Rowan-Robinson et al (2008). Table columns give object number, RA, dec, spectroscopic redshift, 
spectroscopic type, $log_{10}$ bolometric luminosity in cirrus component
(from template fit), bolometric luminosity in starburst component (M82 
type unless indicated in bracket as Arp 220 type), bolometric luminosity 
in optical component (from template fit), optical template type, 
visual extinction in magnitudes.}
\begin{tabular}{llllllllll}
\hline
\hline
object no. & RA & dec & $z_{spec}$& spec. type  & $L_{cirr}$ & $L_{sb}$ &  $L_{opt}$ & type & $A_V$\\
&&&&&&&&&\\
\hline
\hline
1 & 242.67188 & 54.17559 & 0.227 &LINER/SF &  & 11.58 (A220) & 10.98 & Scd & 0.3\\
2 & 242.61204 & 54.177194 & 0.336 & Composite& 10.72 & 10.67 & 10.80 & Scd & 0.0\\ 
3 & 242.83070 & 54.21578 & 0.069 &SF & 10.15 &  & 10.35 & Scd & 0.3\\
4 & 242.71574 & 54.23111 & 0.063 &SF  & 9.97 &  & 10.77 & Sab & 0.0\\
5 & 242.82278 & 54.27474 & 0.143 &SF & 11.02 &  &  10.92 & E &  0.0\\
6 & 242.82320 & 54.27925 & 0.144 & SF & 10.22 &  & 10.72 & sb & 1.3 \\
7 & 242.77863 & 54.31121 & 0.315 &SF  & 10.80 &  & 10.48 & Scd & 0.2\\
8 & 242.44196 & 54.357222 & 0.066 &SF & 10.15 & & 10.35 & sb & 0.0\\
9 & 242.76622 & 54.40152 & 0.130 & SF& 9.81 & 9.71 & 10.35 & sb & 0.0\\
10 & 243.08987 & 54.43593 & 0.338 &Composite & 10.41 & 10.36 & 10.72 & Scd & 1.0\\
11 & 243.25875 & 54.479611 & 0.126 & SF& 10.6 & & 10.7 & Scd & 0.3\\
12 & 242.42502 & 54.55011 & 0.212 &NLAGN & 10.09 & 9.61 & 10.27 & Scd & 0.6\\
13 & 242.76559 & 54.72282 & 0.063 &NLAGN & 10.05 & 9.15 & 10.35 & Scd & 0.4\\
\hline
\hline
\end{tabular}
\end{table*}

\label{lastpage}
\end{document}